\documentclass[journal=ancac3,manuscript=article]{achemso}

\usepackage{chemformula} 
\usepackage[T1]{fontenc} 
\DeclareUnicodeCharacter{2212}{\textendash}
\usepackage{siunitx}

\author{Jonathan T. Weber}
\affiliation{Institute of Physics, Carl-von-Ossietzky University of Oldenburg, Germany}
\alsoaffiliation[]{Department of Physics, University of Regensburg, Germany}
\author{Sascha Schäfer}
\affiliation{Institute of Physics, Carl-von-Ossietzky University of Oldenburg, Germany}
\alsoaffiliation[]{Department of Physics, University of Regensburg, Germany}
\alsoaffiliation[]{Regensburg Center for Ultrafast Nanoscopy (RUN) University of Regensburg, Germany}
\email{sascha.schaefer@ur.de}

\title[]{Electron Imaging of Nanoscale Charge Distributions Induced by Femtosecond Light Pulses}

\begin{document}

\section{Abstract}

Surface charging is a phenomenon ubiquitously observable in in-situ transmission electron microscopy of non-conducting specimens as a result of electron beam/sample interactions or optical stimuli and often limits the achievable image stability and spatial or spectral resolution.
Here, we report on the electron-optical imaging of surface charging on a nanostructured surface following femtosecond-multiphoton photoemission. By quantitatively extracting the light-induced electrostatic potential and studying the charging dynamics on the relevant timescales, we gain insights into the details of the multi-photon photoemission process in the presence of a background field. We study the interaction of the charge distribution with the high-energy electron beam and secondary electrons and propose a simple model to describe the interplay of electron- and light-induced processes. In addition, we demonstrate how to mitigate sample charging by simultaneous optical illumination of the sample.

\section{Introduction}
Sample charging in electron microscopy results from a number of interlinked interactions between high-energy electrons and nanoscale specimens, such as electronic excitation, defect generation and the emission of secondary electrons \cite{cazaux_correlations_1995}.

For example, in cryo-electron microscopy the interaction of imaging electrons with accumulated charges in amorphous ice films is often detrimental and poses a limit to the achievable spatial resolution, image stability and image contrast \cite{russo_charge_2018,brink_evaluation_1998,bottcher_electron_1995,russo_microscopic_2018}. 
An ongoing effort is made to quantify and understand the contributing processes in detail \cite{schreiber_temporal_2023,glaeser_specimen_2004,beleggia_local_2016} as well as mitigating the adverse effects of sample charging \cite{berriman_paraxial_2012,park_direct_2015}. 

While in many cases sample charging needs to be minimized, other fields like liquid phase electron microscopy \cite{ross_liquid_2016} often rely heavily on the interaction of the sample with the electron beam, utilizing the charge accumulation for electron-beam-induced fragmentation of precursors \cite{gonzalez-martinez_electron-beam_2016,pyrz_electron_2007} or charging-induced ion transport \cite{chen_electron_2014,jiang_situ_2015,white_charged_2012}. \par 

Despite the distinct underlying mechanisms, sample charging in weakly-conducting specimens is also commonly encountered in photoemission spectroscopy and microscopy approaches \cite{suzer_differential_2003,cazaux_about_2000,mihaychuk_multiphoton_1999}.
Here, light-induced surface charging manifests as a shift and, for inhomogeneous charging, as a broadening in the measured photoelectron spectra as well as a suppression of the total photoelectron yield. In some cases sample charging can be counteracted by the use of an additional low-energy electron beam neutralizing the charge distribution \cite{audi_valence-band_2002,huchital_use_1972}. \par
 
Similarly, in the emerging field of electron microscopy with in-situ optical excitation \cite{eggebrecht_light-induced_2017,berruto_laser-induced_2018,voss_situ_2019,bongiovanni_near-atomic_2023} and ultrafast transmission electron microscopy \cite{feist_ultrafast_2017,nabben_attosecond_2023,feist_nanoscale_2018,cao_femtosecond_2021,
cremons_femtosecond_2016,houdellier_development_2018,vanacore_ultrafast_2019,wang_coherent_2020, bucker_electron_2016,kim_light-induced_2020,zhu_development_2020} sample charging is expected to simultaneously occur from light- and electron-beam driven processes. Fully understanding the effects contributing to charge accumulation in these systems necessitates experiments addressing the sample response to optical- and high-energy-electron stimuli as well as the interplay of these effects. \par 

Here, we report on the light-induced charging of individual gold nanostructures, imaged in transmission electron microscopy (TEM). The induced photovoltages are quantitatively extracted by comparing the defocused experimental electron micrographs to electron-optical image simulations, using a numerically calculated electric potential distribution. The effective non-linearity of the underlying photoemission process is precisely measured using interferometrically stable two-pulse excitation and event-based electron detection, gaining insight into the interplay of light- and electron-beam-induced charging phenomena and their significance for photoemission processes in electron microscopy.

\section{Results and Discussion}

\begin{figure}
  \includegraphics[scale=1]{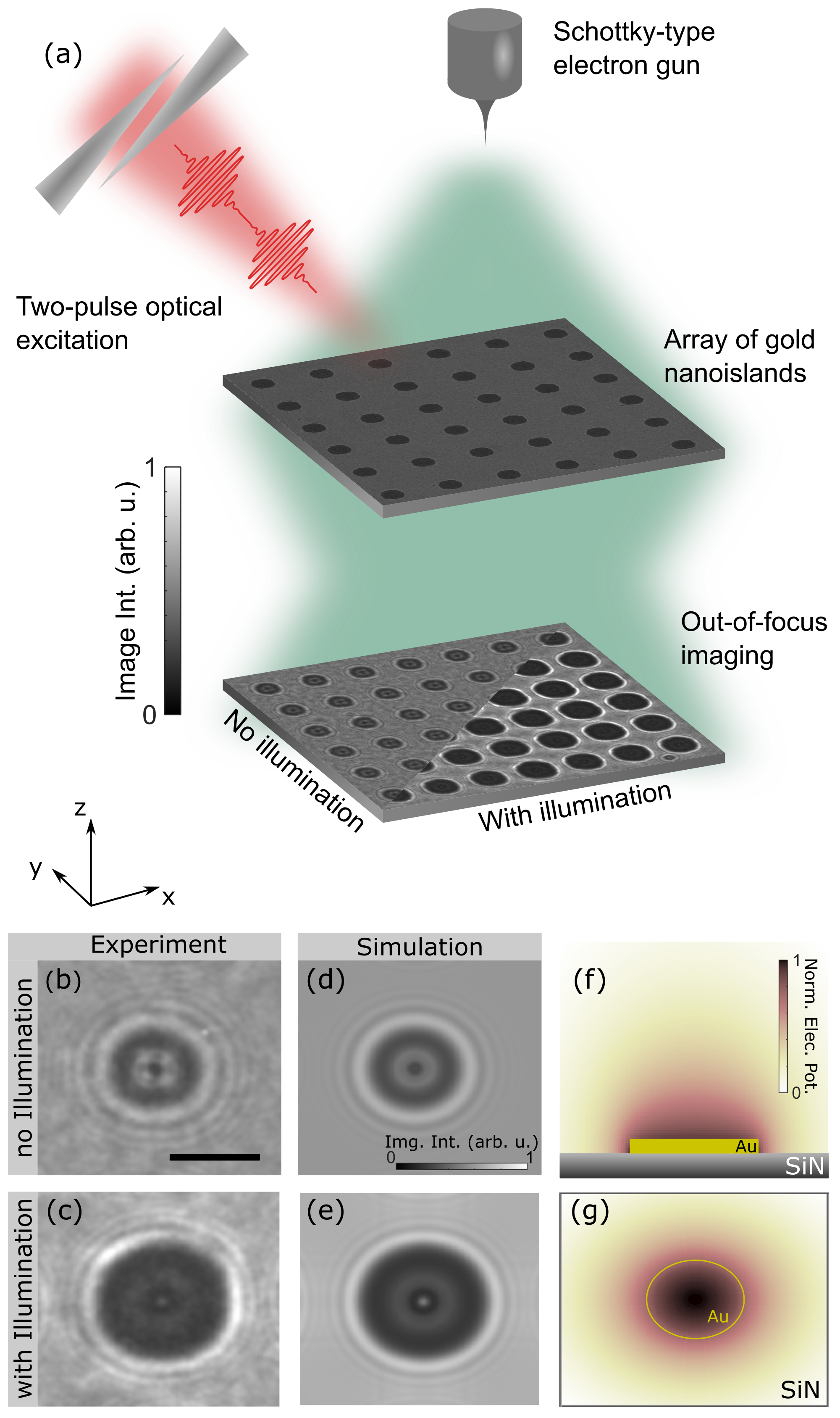}
  \caption{Electron imaging of light-induced charging in a TEM. (a) Scheme of the experimental setup. Optically induced charging of isolated metallic islands in the sample plane results in strong contrast modulations in the image plane under out-of-focus imaging conditions. (b,c) Experimental Lorentz micrographs with/without illumination are compared to image simulations (d,e). Scale bar: 500~nm. The spatial electric potential distribution used in the image simulations is numerically calculated employing a successive over-relaxation approach. The summed distribution is shown in side view (f) and top view (g), with the sample structure indicated by the sketch (not to scale).}
\label{fgr:overview_simulation}
\end{figure}

In order to investigate light-induced charging in electron microscopy, we consider arrays of gold nanoislands on an insulating silicon-nitride membrane as a model system (see Methods for details). Using the Oldenburg ultrafast transmisison electron microscope (UTEM), we illuminate the sample in-situ with femtosecond optical pulses (800-nm central wavelength, 169-fs pulse duration, illumination area widened to 30-µm diameter, 400-kHz repetition rate, p-polarized). For mapping the temporally averaged charge state of individual islands, we employ a continuous electron beam and large electron imaging defoci of $-10.5$~mm. A typical defocused micrograph of a gold island without optical illumination is shown in Fig.~1(b). Upon illumination (1.2-mW average optical power), a drastic change in image contrast occurs (see Fig.~1(c)) resulting in an increase of the apparent nanostructure size by a factor of 1.4 and a change in the electron interference pattern of the nanodisc. Different islands within the illuminated part of the array show comparable light-induced contrast changes with only minor variations, as shown in Fig.~1(a) (lower micrograph).

Using an electron-optical simulation \cite{zuo_advanced_2017}, the image contrast can be quantitatively reproduced considering an optically-induced positive charging of the gold island to a photopotential of +3.9~V, corresponding to a charge depletion by about 500 electrons per island (Fig.~1(d,e)).
The electrostatic potential distribution surrounding the metallic island was determined by numerically solving the electrostatic Laplace equation, using a three-dimensional successive over-relaxation method \cite{hansen_numerical_1992} and cross-checked with a commercial finite-element simulation software (Fig.~1(f) and 1(g)). 
The electric potential imprints a phase shift onto the imaging electron beam \cite{aharonov_significance_1959} which translates into an image contrast in out-of-focus conditions \cite{zuo_advanced_2017}. 
For a quantitative determination of the induced photovoltage with varying optical excitation, the image intensity is fitted by minimizing the squared differences between experimental and simulated image intensities, with the light-induced voltage $U_{\mathrm{PV}}$ on the metallic islands as the only free fitting parameter. We attribute the observed charging to a multi-photon photoemission process facilitated by the high intensities in the femtosecond light pulses (estimated peak intensity of 3.1~GW/$\mathrm{cm}^2$). We note that our experimental conditions are tuned to achieve a long life-time of the charge depleted state, so that the final state after femtosecond charging can be studied with a continuous electron beam. 

In order to further characterize the multi-photon photoemission process, we conducted experiments with phase-stable pairs of collinear optical pulses with adjustable delay. For this purpose a birefringent common-path interferometer was introduced into the optical beam path, similar to a translating-wedge-based identical pulses encoding system (TWINS) \cite{brida_phase-locked_2012}. For pulse delays smaller than the temporal pulse widths, the interferometer modulates the overall optical power due to the interference of both pulses. The modulation period is given by the optical period of the light pulses. Larger delays can be used to investigate potential non-instantaneous light-induced dynamics. \par 

Experimentally, we observe that the defocused micrographs strongly depend on the pulse delay. In Fig.~2(a), image intensity profiles across a single disc are shown for varying pulse delays close to zero. The corresponding recorded optical power (blue line) and extracted photovoltage (red circles) are displayed in Fig.~2(b) (see also Supporting Materials M1). Whereas the optical power exhibits a simple harmonic dependency with the pulse delay as expected, both the experimental profile widths and the extracted photovoltage show a more complex behavior. We accumulate the data from the four optical cycles shown in Fig.~2(b) and plot the photovoltage depending on the light intensity (Fig.~2(c)), confirming that for these delays the photovoltage is given as function of the optical power. 
In a logarithmic plot, the power scaling of the photovoltage at low fluence shows a slope of 6.5, which is higher than the expected value of 4, given the photon energy of 1.55~eV and gold's workfunction of about 5.3~eV \cite{sachtler_work_1966}. At higher light intensities, the photovoltage saturates. A potential explanation could be space charge effects within a photoemitted electron cloud \cite{riffe_femtosecond_1993,bach_coulomb_2019,van_oudheusden_electron_2007}. However, as detailed below, for our system this behavior is linked to the accumulative charging of the sample over successive light pulses.

\begin{figure}
  \includegraphics[scale=1]{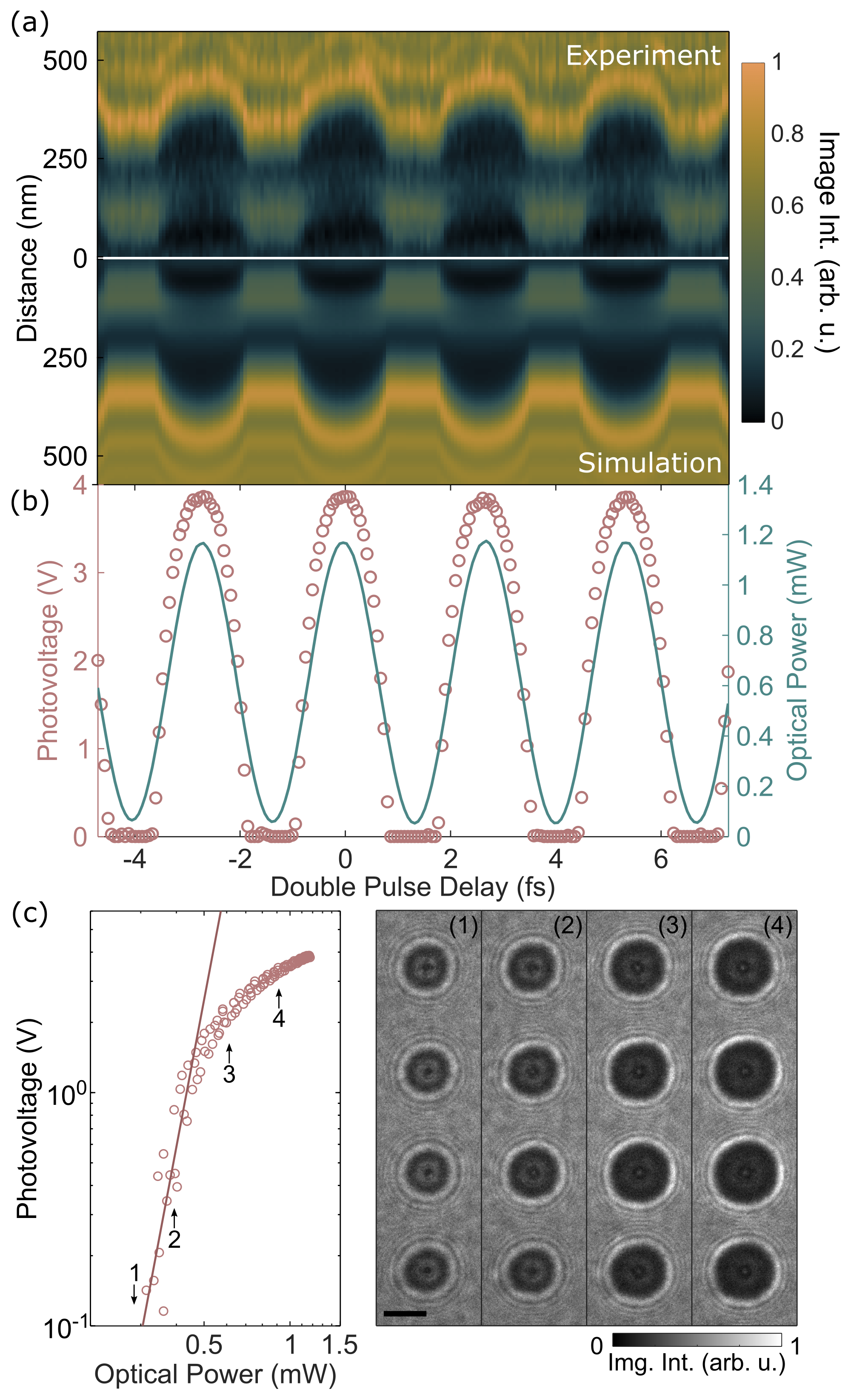}
  \caption{Interferometric two-pulse optical excitation and quantitative determination of the photovoltage. (a) Image intensity profiles through the center of a metallic island, averaged over 3 pixels, for a respective optical power from experimental micrographs (top) and simulated image intensities (bottom).
(b) Incident optical power (blue line) during an interferometric two-pulse excitation measurement in the vicinity of zero pulse-to-pulse delay. The optically induced voltage on a metallic island (pink circles) is determined by non-linear least squares fitting of the image simulation to experimental micrographs with the photovoltage as the only free parameter. 
(c) Light-induced voltage on a metallic island as a function of the optical power. The straight line represents a slope of 6.5 in the logarithmic plot. The right panel show experimental micrographs of a column of gold islands at the indicated optical power. Scale bar: 500~nm.}
  \label{fgr:twins}
\end{figure}

We further investigated the induced photovoltage for larger pulse delays (see Fig.~3(a)). The optical power variation traces the field autocorrelation of the optical excitation pulse (spectral width of about 6.8~nm). 
Due to the above-mentioned nonlinear intensity dependence, the photovoltage shows a distinctly different behavior compared to the autocorrelation but without clear signature of a delayed sample response, for example due to a hot electron gas.

In order to understand the apparent saturation of photoemission efficiency at higher fluences, we apply an electron microscopy approach with high temporal resolution and investigated the accumulative charging over successive optical pulses by using an event-based electron detector based on a TimePix3 chip architecture \cite{poikela_timepix3_2014,schroder_improving_2023,van_schayck_integration_2023}. Using a Pockels cell, we precisely chop the optical excitation at a frequency of 5~Hz (50\% duty cycle) and collect an 8-s long electron event stream on the TimePix detector. The events are sorted into bins of 500-$\mu$s width according to their relative delay to the 5-Hz control signal. The photovoltages extracted from these reconstructed micrographs (see also Supporting Materials M2 and M3 for assembled movies) are shown in Fig.~3(d) for an electron beam dose rate of 0.012~electron/$(\mathrm{nm}^2 \mathrm{s})$ (pink circles) and 0.043~electron/$(\mathrm{nm}^2 \mathrm{s})$ (yellow circles), respectively. 

For delays from 0 to 100~ms no illumination of the sample occurs and a decharging of the metallic islands is observable. This process is governed by the emission of secondary electrons in the vicinity of the charged islands by the incident high-energy electron beam, which neutralize the positively charged metallic nanostructures \cite{russo_charge_2018}. Generally, the number of emitted secondary electrons depend on the electron dose and the substrate material. In our case, part of the silicon frame which holds the silicon nitride membrane is iluminated by the electron beam, resulting in a higher secondary electron yield.
Consequently, different decharging rates of 8~V/s and 21~V/s are observed for the different electron beam dose rates, respectively. A further contributing charging mechanism might be an electron-beam-induced increase in the electrical conductivity of the silicon nitride substrate. \par 

Upon re-illumination of the sample at a delay of 100~ms, a fast increase of the photovoltage is observable, with a time constant of about 2~ms. Subsequently, after the accumulated effect of about $10^3$ optical pulses, the measured photovoltage remains at a constant value $U_{\mathrm{PV}}$ for the remaining duration of optical excitation. 

\begin{figure}
  \includegraphics[scale=1]{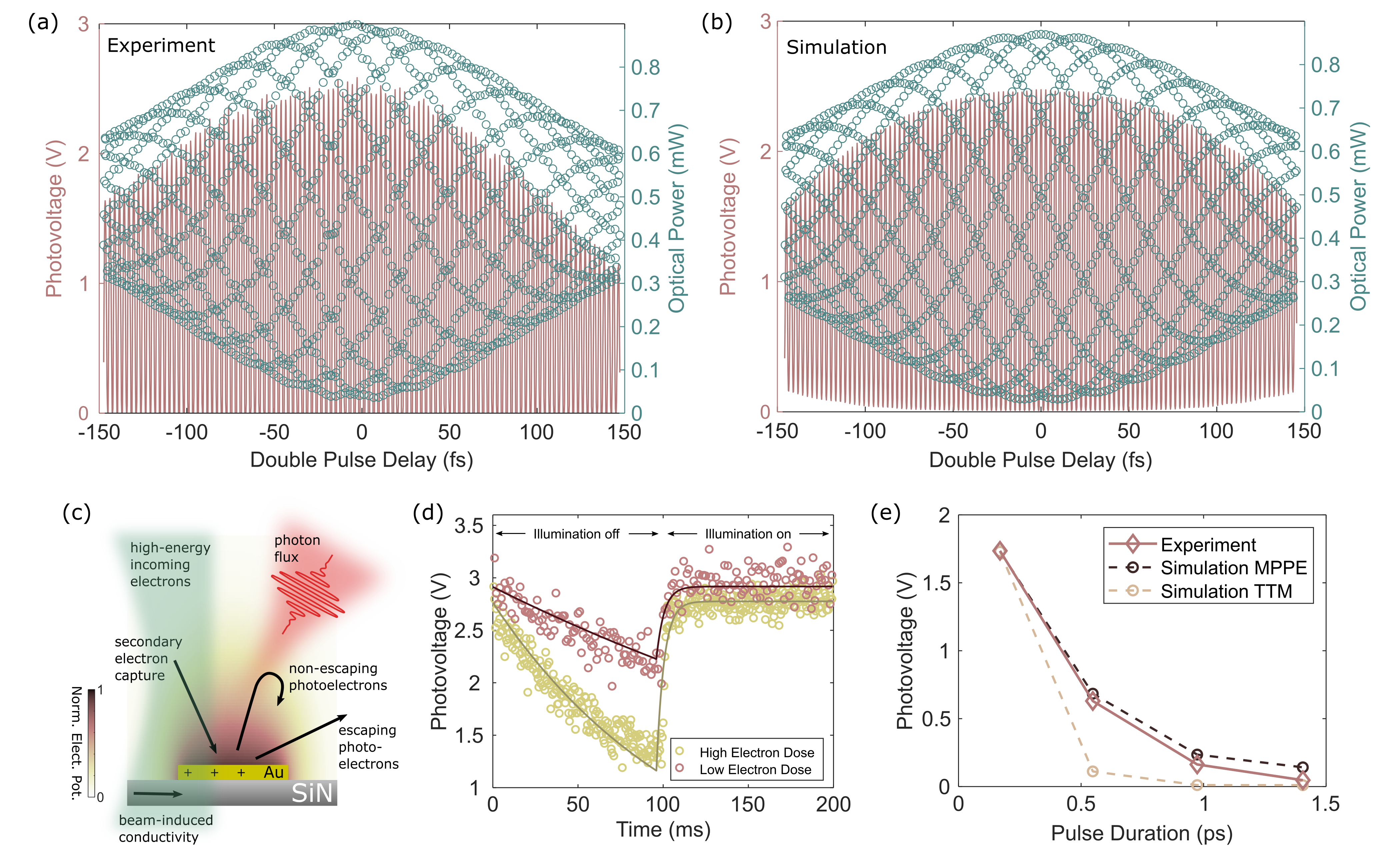}
 \caption{Dynamics of the charging/decharging process and interplay of light- and electron-beam-induced phenomena. 
(a) Photovoltage (pink line) during interferometric double-pulse optical excitation for pulse delays up to $\pm$ 150~fs. Measurement of the optical power (blue circles) corresponds to the field autocorrelation function.
(b) Simulation of the photovoltage for varying pulse-to-pulse delays of the optical excitation in an interferometric measurement scheme.
(c) Overview of the electron-beam- and light-induced processes contributing to the charging/decharging of individual metallic islands. 
(d) Charging/Decharging cycle measured with an event-based electron detector for an electron dose rate of 0.012~electron/$(\mathrm{nm}^2 \mathrm{s})$ (pink circles) and 0.043~electron/$(\mathrm{nm}^2 \mathrm{s})$ (yellow circles). The dynamics are modeled with a rate equation (dark red and yellow lines) including optical- and electron-beam-induced contributions to the charge state of the islands (see text for details).
(e) Experimentally measured photovoltages for optical excitation with varying pulse lengths and comparison to the simulated results considering a multiphoton-photoemission process and a thermionic emission process. Simulations are adjusted to replicate the experimental photovoltage value for unstretched light pulses.
}
  \label{fgr:dynamics}
\end{figure}

The observed behavior can be explained within the framework of photoemission in a background electric field. Photoemitted electrons have to overcome the electrostatic potential that surrounds the charged islands. As a consequence, electrons with insufficient initial energy from the multi-photon absorption are unable to escape the potential well and instead fall back to the surface of the metallic island, not contributing to a further charging of the island. At a saturated photovoltage (at delays larger than 100~ms), only few photoelectrons escape the Coulomb potential around the islands which balance the electron-beam-induced decharging, thereby maintaining an equilibrium potential state of the island. Supporting this picture, TimePix-based recordings with nanosecond time bins (see Supporting Material M4) did not show any apparent contrast changes in-between optical pulses, highlighting that in the saturated state, only minimal charging occurs. The contributing processes are sketched in Fig.~3(c). \par 

Along these lines also the observed distinctive intensity-dependence, characterized by a consistent decrease in slope as optical power increases, can be explained. We describe the dynamic charging process with a rate equation model, in which the change of the photovoltage $U_{\mathrm{PV}}$ of an island is given by 
\begin{align}
\frac{\mathrm{d}U_{\mathrm{PV}}}{\mathrm{d}t} = - k_1 I_e \sigma_{\mathrm{capture}}(U_{\mathrm{PV}}) + k_2 I_p^n \sigma_{\mathrm{escape}}(U_{\mathrm{PV}}) + k_3 I_e.
\end{align}
The first part of the expression corresponds to the decharging of the metallic nanoisland due to the capture of secondary electrons induced by the beam current $I_e$. The capture probability $\sigma_{\mathrm{capture}}(U_{\mathrm{PV}})$ depends on the charge state of the islands. For simplicity, we consider a simple relation $\sigma_{\mathrm{capture}}(U_{\mathrm{PV}})=U_{\mathrm{PV}}$ which results in an exponential decay of the island's charge state. 
The second term describes the charging of the islands due to photoemission and depends on the optical intensity $I_p$, the effective non-linearity of the photoemission process $n$ and the probability $\sigma_{\mathrm{escape}}(U_{\mathrm{PV}})$ that a photoemitted electron escapes the electrostatic potential of the charged island.
The third term represents the contribution of the electron-beam-induced positive charging to the overall dynamics.
$k_1$, $k_2$ and $k_3$ are rate constants of the involved processes. 
The escape probability can be connected to the photoelectron energy distribution $g(E)$, yielding 
\begin{align}
\sigma_{\mathrm{escape}}(U_{\mathrm{PV}})=\int_{eU_{\mathrm{PV}}}^\infty g(E) dE.
\end{align} 
In general, $g(E)$ will change with the optical intensity due to the opening of multiphoton emission channels with higher nonlinearities at increased intensities. 
By using a constant photoelectron energy distribution with an upper limit of 2.99~eV, we already obtain a good fit to the experimental time-resolved photovoltage traces, as shown in Fig.~3(d).
With the same set of fit parameters also the absolute value of the photovoltage induced by double-pulse excitation can be well reproduced (Fig.~3(a,b)), despite the different electron beam dose rates and temporal optical pulse shapes employed for these experiments. Minor misalignments in the common path interferometer lead to a non-vanishing fluence for destructive interference around zero delay. This is taken into account in the simulations by assigning slightly different amplitudes to the electric fields of the two pulse copies. \par
To clarify the characteristics of the photoemission process we further conducted measurements with varying optical pulse lengths (see Fig.~3(e)). We incorporated dense-flint glass cylinders with an effective length of 10~cm (20~cm, 30~cm) into the optical beam path, thus introducing a strong chirp to the pulses, effectively stretching the optical excitation to a duration of 0.55~ps (0.97~ps, 1.4~ps). The comparison of the experimental values (pink circles) to the photovoltages expected from a pure multi-photon photoemission process in a background field (black cirlces) and a thermionic emission process in a background field (yellow circles), calculated by utilizing a simple two-temperature model, suggests, that the multi-photon pathway is dominating at the employed excitation parameters. \par

\begin{figure}
  \includegraphics[scale=1]{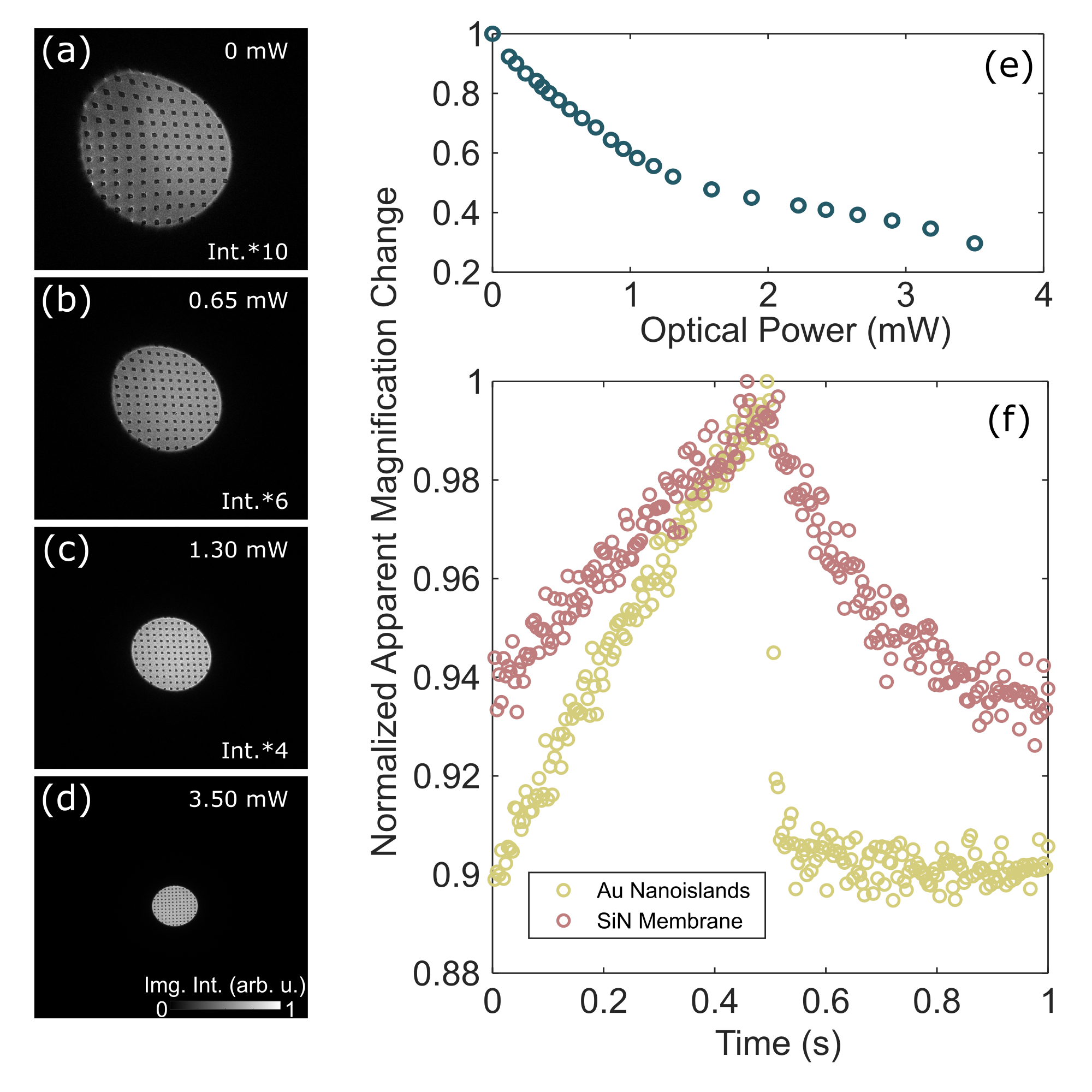}
 \caption{Mitigating electron-beam-induced charging by optical illumination. (a-d) Experimental Lorentz micrographs showing a strong electron-beam-induced change of the apparent magnification and image distortions (electron dose rate: 0.02~electron/$(\mathrm{nm}^2 \mathrm{s})$, defocus value: -10~mm). The sample is illuminated with femtosecond light pulses with the indicated average optical power. The image intensity is multiplied with the indicated factor to increase contrast. The nanoisland periodicity is 1.5~$\mu$m. 
(e) Change of the apparent magnification for an array of illuminated gold nanoislands as a function of average optical power. The apparent magnification is measured as the gaussian width of the electron spot along one dimension.
(f) Change of the apparent magnification measured with an event-based electron detector while chopping the optical excitation (50\% duty cycle, 1 Hz). The decharging dynamics are investigated on an array of gold nanoislands (optical power: 0.8~mW, yellow circles) and an empty silicon nitride membrane (optical power: 2.3~mW, pink circles). Both measurements were conducted with an electron dose rate of 0.012~electron/$(\mathrm{nm}^2 \mathrm{s})$ and a defocus value of -20~mm. We note that the photoinduced magnification changes sensitively depend on the illumination conditions.
}
  \label{fgr:4}
\end{figure}

Whereas - so-far - we focussed on the quantitative description of the light-induced charging and photoemission characteristics, we note that light-driven processes can also be utilized to compensate electron-beam-induced charge accumulation. At illumination conditions for which the primary beam is not impinging on the silicon support frame of the membrane, secondary electron emission is minimized, and other decharging mechanisms may become relevant. In Fig.~4(a-d), gold islands are imaged under such conditions and at different light intensities. Without optical illumination (Fig.~4a), the image of the island array is strongly distorted and changed in its effective magnifcation by a factor of 3.5 due to an electron-beam-induced local charge accumulation and thereby the formation of an electrostatic lens. As compared to the laser-induced charging experiments, less contrast is seen around each nanoisland, putatively due to a more homogeneous charge distribution in the electron-beam-induced case. We observe a reduction of the apparent magnification upon illuminating the charged area with fs-light pulses, indicating a neutralization of the accumulated electron-beam-induced charge (see Fig.~4(a-d)). The decharging effect becomes more pronounced for higher optical powers (see Fig.~4(e), optical spot diameter 10-15$\mu$m) and occurs only when the spatial overlap between the electron beam and the optical focus is maintained. \\
We investigate the dynamics of charge buildup and light-induced neutralization using the event-based electron detection scheme, while chopping the optical excitation at a frequency of 1~Hz (50\% duty cycle), on an array of gold nanoislands as described above and an empty silicon nitride membrane. For delays between 0 and 500~ms no optical illumination occurs and in both samples an increase of apparent magnification due to electron-beam-induced charge accumulation is observable. Re-illumination at a delay of 500~ms results in a decrease of apparent magnification. For the silicon nitride membrane (without gold islands) we observe a progressive decharging for the remainder of the illumination period, just starting to saturate at a delay of 1~s. If additionally gold nanoislands are present in the area under investigation, we observe a step-like compensation of the accumulated charge, reaching an equilibrium state after a few milliseconds. \\
Optical illumination has been reported to reduce surface charging effects in electron microscopy \cite{seniutinas_ultraviolet-photoelectric_2016}, using photon energies close to the materials workfunction by detrapping accumulated charges. As a direct excitation of surface defects is expected to be largely suppressed at the wavelength employed in our experiments, we hypothesize that instead the observed behavior could be attributed to an optical-pumping-induced, locally increased sample temperature. The enhanced electrical conductivity would shift the equilibrium point of electron-beam-induced charge accumulation to smaller values. Additionally, internal photoemission processes might contribute to the observed behavior.

We note that optical illumination serves as an effective method for mitigating the detrimental impacts of sample charging in electron microscopy, reducing electron-beam-induced lensing effects by a factor of up to 3 in our experiments. Beyond this application, the strong image contrast modulations upon optical illumination through either charging of individual gold nanoislands or decharging of the sample, provides a viable tool for finding the spatial overlap between an electron beam and optical foci on the sample, necessary in optical in-situ TEM and in UTEM approaches - a usually very time-consuming task requiring meticulous alignment. Furthermore, this technique offers an easy approach to estimate the size of the illuminated area on the sample (see Supporting Materials Figure S1 for electron micrograph of an inhomogeneously illuminated array of square-shaped nanoislands).

\section{Conclusion}
In conclusion, we presented the electron-optical imaging of light-induced charge distributions on a nanostructured surface. We quantitatively determined the photovoltage by reproducing the experimental micrographs with electron-optical image simulations using a numerically calculated electrostatic potential distribution. By utilizing interferometric two-pulse excitation measurements and event-based electron recording, we could identify the underlying process as a multi-photon photoemission process in a background electric field and in the presence of low-energy secondary electrons.
We modelled the charging dynamics with a rate equation and quantified the contributions of light- and electron-beam-induced effects. With the same set of parameters, we were able to quantitatively reproduce the observed photovoltages for different electron beam doses, optical powers and effective optical pulse lengths, highlighting the quality of our model. In the future, our findings may help to disentangle the various charging-related phenomena and enable a more precise and controlled characterization of nanoscale materials and devices. First results on light-induced decharging processes show potential to mitigate adverse effects of charging dynamics in high-resolution electron microscopy.

\section{Acknowledgment}
The authors acknowledge financial support by the Volkswagen Foundation as part of the Lichtenberg Professorship "Ultrafast nanoscale dynamics probed by time resolved electron imaging". The authors also thank the German Science Foundation for the funding of the ultrafast transmission electron microscope (INST 184/211 1 FUGG) and the electron beam lithography instrument (INST 184/107-1 FUGG).

\section{Methods}

\noindent \textbf{Specimen Preparation:} \\
The investigated samples consist of an array of disc-shaped Au islands with a diameter of 500~nm and an inter-island spacing of 1~$\mathrm{\mu}$m. Using a lift off process, the specimens were prepared on 50-nm thick silicon nitride membranes (PELCO) as a substrate. A mask is patterned by electron beam lithography into a poly(methyl methacrylate) (PMMA) resist with subsequent development. Using electron beam vapor deposition (base pressure of $10^{-7}$~mbar), the sample was coated with a 3-nm chromium adhesion layer, followed by a 17-nm layer of gold. The final structure thickness was confirmed by atomic force microscopy measurements. \\

\noindent \textbf{Electron Microscopy:} \\
Electron micrographs were recorded with the Oldenburg ultrafast transmission electron microscope which is based on a JEOL JEM-F200 instrument (200-keV electron energy, Schottky-type electron gun). The microscope was operated in the low magnification mode with the objective lens turned off and a defocus of -10.5~mm was chosen, unless stated otherwise. For electron illumination, a 100-$\mathrm{\mu}$m diameter condenser aperture and a spot size of five was used. Micrographs were acquired with a complementary metal-oxide-semiconductor (CMOS) detector (TVIPS TemCam-XF416R, 4069x4069 pixels, 15.5-µm pixel size). Matlab was used for all further evaluation steps, which included binning (4x4 pixels) and Gauss filtering (Standard deviation of 2D Gaussian smoothing kernel: 2) and the analysis of the photovoltage. For time-resolved measurements, we utilized a TimePix3 detector (Cheetah T3, Amsterdam Scientific Instruments) which is an event-based electron detector with a nominal time bin width of 1.6~ns. \\

\noindent \textbf{Optical Setup:} \\
For triggering multiphoton-photoemission from the gold nanoislands, we used optical pulses from a collinear optical parametric amplifier (OPA, Orpheus HP, Light Conversion) seeded and pumped by an amplified Yb-doped potassium gadolinium tungstate (KGW) femtosecond laser system (Carbide, Light Conversion). Optical pulses were characterized by a self-built frequency-resolved optical gating setup (FROG).
Femtosecond light pulses are focused on the sample using an incoupling unit installed on a flange located at the height of the TEM pole piece. The incoupling unit consists of a vacuum viewport and a focusing lens (focal length: 50~mm, diameter: 0.5'') which is mounted on three piezo stages, enabling high-precision scanning of the optical focus over the sample, with an incident angle of about 57° relative to the electron beam. An active beam stabilization system (Aligna, TEM Messtechnik) is utilized, to accommodate relative movements of the optical laser system and the TEM column, each supported on individual vibration damping systems. \\
For the generation of interferometrically stable optical pulse pairs with adjustable delay, we used a birefringent common-path interferometer \cite{brida_phase-locked_2012}. Specifically, a half-wave plate is used to polarize the light at 45° relative to the fast axis of a planar $\alpha$-BBO crystal (thickness: 4~mm). The fast axis of the following pair of $\alpha$-BBO wedges (length: 50~mm, opening angle: $7$°) is rotated by $90$° with respect to that of the planar $\alpha$-BBO crystal. The last element of the interferometer is a polarization filter, with the transmission axis at a 45°~angle with respect to the fast axes of the planar and the wedged $\alpha$-BBO crystals. \\

\noindent \textbf{Supporting Information:} \\
Additional details on the data analysis and computational methods, including a micrograph showing an inhomogeneously excited array of nanoislands (PDF) \\
M1 Assembled video of the data presented in Fig.~2 (AVI) \\
M2 Assembled video of the data presented in Fig.~3, high-dose electron beam (AVI) \\
M3 Assembled video of the data presented in Fig.~3, low-dose electron beam (AVI)\\
M4 In-situ observation of nanoislands, optically excited with 400-kHz repetition rate (AVI) \\

\newpage

\bibliography{Charging_Paper_Bib.bib}

\newpage
\setcounter{equation}{0}
\setcounter{figure}{0}
\setcounter{table}{0}
\setcounter{page}{1}
\makeatletter
\renewcommand{\theequation}{S\arabic{equation}}
\renewcommand{\thefigure}{S\arabic{figure}}

\noindent\textbf{\large{Supporting Information for: \\
Electron Imaging of Nanoscale Charge Distributions Induced \\ 
by Femtosecond Light Pulses}}

\vspace{2cm}
\noindent \textbf{This supporting information includes:} \\ 
Additional Information on the image contrast simulation, the electrostatic potential model and the image fit procedure. \\
Figure S1 \\
Captions for Movies 1-4 \\
\textbf{Other supporting materials for this manuscript:} \\
Movies 1-4

\subsection{Image Intensity Calculations}
The wavefunction $\Psi$ of the electron wave after passing the sample is spatially phase-modulated by the charged nanoislands and expressed as $\Psi(\vec{r}) = A(\vec{r}) e ^{i\Phi(\vec{r})}$, with $A(\vec{r})$ and $\Phi(\vec{r})$ being the amplitude and phase imparted on the incident electron plane wave $\Psi_0$. The vector $\vec{r}$ denotes the position in planes perpendicular to the electron trajectory which is chosen to be the z-direction. For numerical simulations, we discretize $\Psi (\vec{r})$ on a 1024x1024 grid, with an effective pixel spacing of 6.6~nm. The phase shift imprinted on the incident electron wavefront is given by the Aharonov-Bohm phase$^{1}$ in the weak-deflection limit

\begin{equation}
  \Phi (\vec{r}) = \frac{e}{\hbar v^*} \int V \left( \vec{r} ,z \right) \mathrm{d}z. 
  \label{eqn:aharonov-bohm1}
\end{equation}
Here, $V \left( \vec{r} ,z \right)$ is the electrostatic potential, $e$ is the electron charge, $\hbar$ the reduced Planck constant and $v^*$ is the relativistic velocity of the electrons used for imaging. Any magnetic contributions are disregarded. 
Eq.~\ref{eqn:aharonov-bohm1} is used to calculate the phase shift introduced by the mean-inner potential $V_\textrm{MIP}=28 V$ of the thin gold islands,$^2$ as well as for the additional phase-shift due to the electrostatic potential induced by optical excitation. For the mean inner potential, Eq.~\ref{eqn:aharonov-bohm1} simplifies to $\Phi = e/(\hbar v^*) V_\textrm{MIP}t_{\mathrm{Au}}$ with $t_{\mathrm{Au}}$ as the thickness of the gold film. \\
For the contribution of the charging-related phase shift the integral is carried out along the z-axis over the three-dimensional electrostatic potential around a charged metallic island and the potential within the island itself. 
Details on the numerical calculation of the potential distribution are found in the following section. \par 
From the wavefunction $\Psi (\vec{r})$ the resulting image $I (\vec{r})$, produced by an electron-optical system in Lorentz-mode can be calculated by
 \begin{equation}
   I (\vec{r})= |\mathcal{F}^{-1} \left[ T(\vec{q}_{\perp}) \mathcal{F} \left(\Phi (\vec{r}) \right) \right] |^2 .\label{eqn:contrast-transfer}
\end{equation}
Here, $T(\vec{q}_{\perp})$ is the contrast transfer function defined in reciprocal space, with $\vec{q}_{\perp}$ as a wavevector in the plane perpendicular to the electron trajectory, and $\mathcal{F}$ as the Fourier Transform in the transverse plane.
The contrast transfer function can be written as 
\begin{align}
T(\vec{q}_{\perp}) &= e^{-i\chi(\vec{q}_{\perp})}  e^{-g(\vec{q}_{\perp})} 
\end{align}
in which $ e^{-i\chi(\vec{q}_{\perp})}$ is the phase-transfer function and $ e^{-g(\vec{q}_{\perp})}$ is a damping envelope incorporating a finite spatial coherence (temporal coherence effects are neglected).
The parameters $\chi(\vec{q}_{\perp})$ and $g(\vec{q}_{\perp})$ are given by

\begin{align}
\chi (\vec{q}_{\perp}) &= -\frac{2 \pi}{\lambda} \left( \frac{\Delta f}{2} |\vec{q}_{\perp}|^2 \right) \\
g(\vec{q}_{\perp}) & = \frac{\left( \pi \theta_c \Delta f \right)^2}{\mathrm{ln} 2} |\vec{q}_{\perp}|^2
\end{align}
with $\Delta f$ as the defocus of the imaging system and $\theta_c$ as beam divergence. The influence of spherical lens aberrations can be neglected due to the large defocus.$^3$ Using a defocus of $\Delta f=-10.5$~mm, the observed Fresnel fringes around the gold islands (without illumination) are well reproduced (see Fig. 1(b,d) in the main text). 
No aperture is placed in the back-focal plane of the imaging lens. 
High-angle electron scattering within the gold islands and the small acceptance angle of the imaging system in Lorentz mode result in a decreased image intensity at the island position. We model this effect by considering an additional effective amplitude modulation $\sqrt{I_{mod}}$ for the electron wave components passing a gold island. The value for $I_{mod}$ is extracted from in-focus images. \\

\subsection{Calculation of the Electrostatic Potential}
\label{successive}
For calculating the electrostatic potential distribution $V_{\mathrm{es}}$ around the charged nanoislands, we adopt a numerical solution scheme for the corresponding Laplace equation in cylinder coordinates$^4$
\begin{align}
\bigtriangleup V \left( \rho, \phi, z \right) &= \frac{1}{\rho}\frac{\delta}{\delta \rho}\left( \rho 	\frac{ \delta V}{\delta \rho}\right) + \frac{1}{\rho^2}\frac{\delta^2V}{\delta \phi^2} + \frac{\delta^2 V}{\delta z^2} \\
&= 0.
\end{align}
On the surface of the island the potential is constant. In the region surrounding the island, the electric potential can be approximated by considering a discretized version of the Laplace operator in the $(\rho, z)$-plane, utilizing the cylinder symmetry of the geometry. Thereby, the electrostatic potential $V_{\mathrm{es}}  \left(\rho, z \right)$ at any position outside of the gold nanoislands needs to satisfy the relation

\begin{align}
V_{\mathrm{es}}  \left(\rho, z \right) = &\frac{1}{4} \bigl [V_{\mathrm{es}}(\rho+h,z) + V_{\mathrm{es}}(\rho-h,z) + V_{\mathrm{es}}(\rho,z+h) + V_{\mathrm{es}}(\rho,z-h) \bigr ] \nonumber \\
&+ \frac{h}{\rho} \bigl [ V_{\mathrm{es}}(\rho+\frac{1}{2}h,z) + V_{\mathrm{es}}(\rho-\frac{1}{2}h,z)  \bigr ].
\label{eq:successive_over}
\end{align}

In order to find a function $V_{\mathrm{es}}  \left( \rho, z  \right)$ that satisfies this condition, an initial guess for $V_{\mathrm{es}} $ is placed on a equidistant 1024x1024 grid (effective pixel spacing 6.6~nm). This initial guess is relaxed towards a more accurate solution by changing the entries based on Eq.~\ref{eq:successive_over}. The relaxation process is repeated for several iterations until the desired level of accuracy is achieved, while ensuring that the constraints imposed by the boundary conditions are satisfied. In our case the boundary conditions are a constant potential on the metallic island itself and vanishing potential at the box boundaries.

\subsection{Extracting the light-induced voltage}
To determine the light-induced voltage on the metallic islands, we use electron imaging simulations to reproduce the experimental data. Using a non-linear least squares algorithm, the image intensity $I (\vec{r})$ calculated by Eq.~\ref{eqn:contrast-transfer} is fitted to dark-corrected normalized and drift-corrected experimental micrographs by minimizing the squared intensity differences summed over all pixels. Here, the light-induced voltage $U_{\mathrm{PV}}$ on the metallic island is the only free fitting parameter. \\
As all mathematical operations in Eq.~\ref{eq:successive_over} are linear, the potential distribution does not have to be calculated for each value of $U_{\mathrm{PV}}$ in the fitting process. \\

\begin{figure}[H]
  \includegraphics[scale=1]{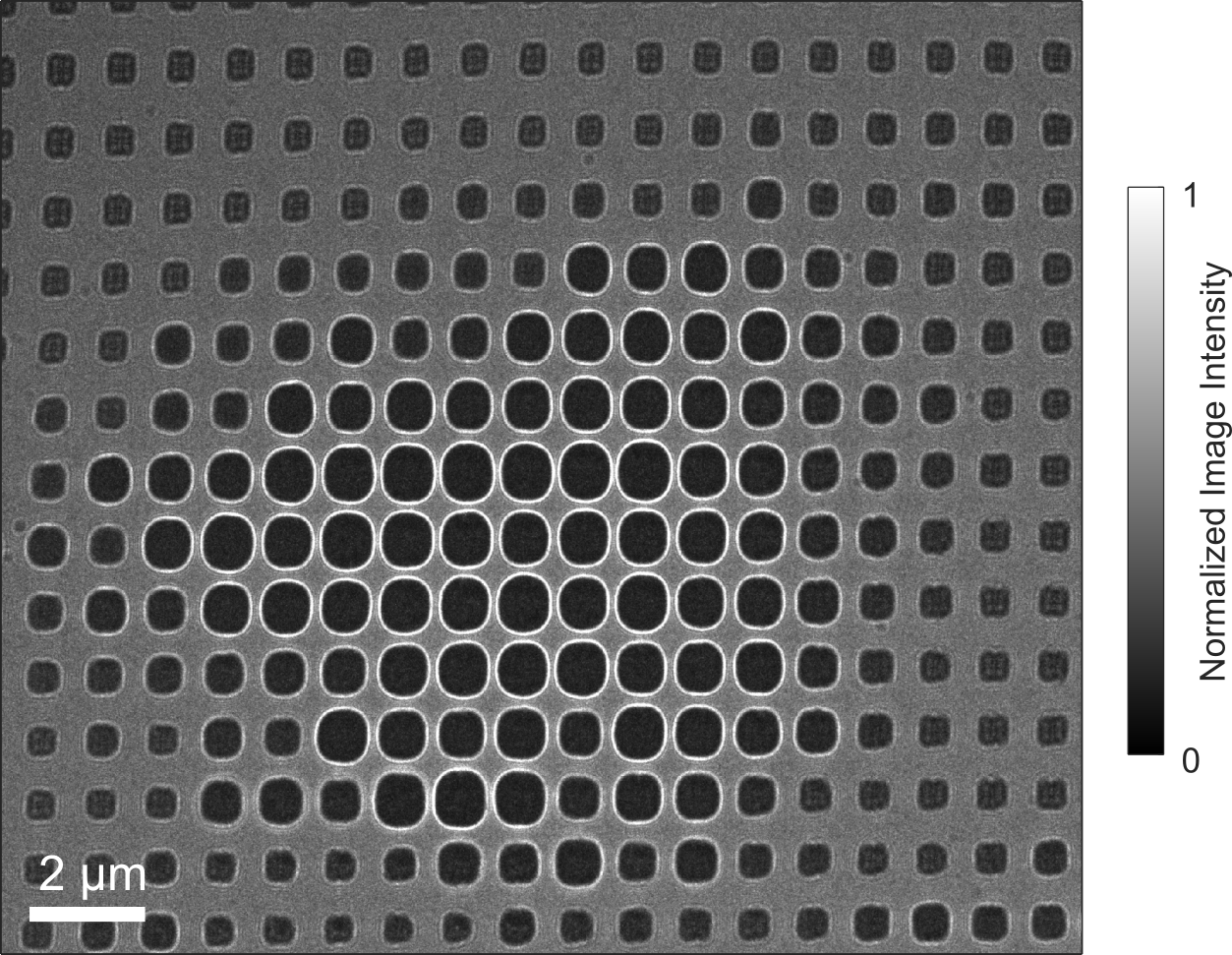}
 \caption{Defocused electron micrograph of an array of square-shaped gold nanoislands on a silicon nitride substrate. The center of the array is illuminated with high intensity light pulses (800-nm central wavelength, 190-fs pulse duration, 400-kHz repetition rate, p-polarized, average power: 1.2~mW). Different from the results in the main text, optical pulses were tightly focused to an area with 10~-~15~µm diameter.}
  \label{fgr:findingOverlap}
\end{figure}

\newpage

\noindent \textbf{Caption for Movie 1:} \\
Assembled video showing defocused electron micrographs of optically excited gold nanoislands while scanning the pulse-to-pulse delay using a common-path birefringent interferometer. \\
\textbf{Caption for Movie 2:} \\
Assembled video of optically excited gold nanoislands recorded with an event-based electron detector based on a TimePix3 chip architecture. The video shows the reconstructed electron micrographs while chopping the optical excitation at a frequency of 5~Hz with a 50\% duty cycle. We used an electron dose rate of 0.043 electron/$(\mathrm{nm}^2 \mathrm{s})$.\\
\textbf{Caption for Movie 3:} \\
Assembled video of optically excited gold nanoislands recorded with an event-based electron detector based on a TimePix3 chip architecture. The video shows the reconstructed electron micrographs while chopping the optical excitation at a frequency of 5~Hz with a 50\% duty cycle. We used an electron dose rate of 0.012 electron/$(\mathrm{nm}^2 \mathrm{s})$. \\
\textbf{Caption for Movie 4:} \\
Assembled video of optically excited gold nanoislands recorded with an event-based electron detector based on a TimePix3 chip architecture. The video shows the reconstructed electron micrographs between consecutive light pulses with a repetition rate of 400~kHz. No change in image contrast is observable.

\newpage
\noindent \textbf{\Large{References}}
\begin{enumerate}
\item Aharonov, Y.; Bohm, D. Significance of Electromagnetic Potentials in the Quantum Theory. \textit{Phys. Rev.} \textbf{1959}, \textit{115}, 485-491.
\item Sanchez, A.; Ochando, M. A. Calculation of the Mean Inner Potential. \textit{J. Phys. C: Solid State Phys.} \textbf{1985}, \textit{18}, 33-41.
\item De Graef, M. \textit{Experimental Methods in the Physical Sciences}; Elsevier, 2001; Vol. 36; pp 27-67.
\item Hansen, P. B. Numerical Solution of Laplace's Equation. \textit{Electrical Engineering and Computer Science - Technical Reports} \textbf{1992}, \textit{168}.
\end{enumerate}

\end{document}